# Powder Lot Variations: A Case Study with Varget – Hodgdon Extreme

Elya Courtney and Michael Courtney

BTG Research, 9574 Simon Lebleu Road, Lake Charles, LA, 70607
Michael_Courtney@alum.mit.edu

**Abstract**
Small arms propellant distributor Hodgdon claims that rifle powders in its Extreme line have small velocity variations with both temperature changes and lot number. This paper reports on the variations in average velocity of four different lots of Hodgdon Extreme Varget tested in two .223 Remington loads. Compared to the lot with the slowest average velocity, the other three lots of powder had higher average velocities ranging from 23.4 ft/s faster up to 45.6 ft/s faster with a 69 grain Nosler Custom Competition bullet and from 7.9 ft/s faster to 15.3 ft/s with the 53 grain Hornady VMAX. The mean velocity differences between lots are slightly correlated between the two loads with a correlation coefficient of 0.54. This correlation suggests that factors other than lot to lot variations contribute significantly to the measured velocity variations.  Unlike the much larger lot to lot variations that were reported previously for H4831, lot to lot variations in velocity for Varget seem consistent with Hodgdon's marketing claims.

**Key Words:** bullet velocity, internal ballistics, variation, powder lot, nitrocellulose

**Introduction**
The Hodgdon Extreme line of powders has long been a top choice of match shooters and long range hunters, in large part due to its marketing claims regarding smaller temperature variations than other brands and small lot to lot performance variations. (Hodgdon 2012a) We were somewhat surprised that our earlier study (Courtney and Courtney 2013) documented lot to lot velocity variations above 100 ft/s for H4831 in both .25-06 and .300 Win Mag.  We received some feedback from Hodgdon about that earlier report, and carefully reviewed our methods, data, and conclusions in light of the feedback.

The information on decoding Hodgdon's lot numbers revealed that the earlier study only included four truly unique lots (different factory runs), and that the lots we had described as C and E were both the same factory lots, but packaged on different dates and acquired through different channels, as were the lots B and D.  This was fascinating because it allowed us to determine that the differences due to unique packaging dates, shipping and handling histories, and storage histories prior to purchase only contributed to velocity variations in the 12 to 17 ft/s range; whereas, the largest differences (over 100 ft/s) were attributable to different factory runs.

After completing the first study in a longtime favorite slow powder (H4831) that we've used for years in overbore cartridges, we were eager to turn our attention to a faster burning powder that has long been our "go to" powder in .223 Remington and .308 Winchester. Varget is a popular powder for many reloading applications in a wide variety of cartridges, so it was a natural choice.  However, supply and demand issues have recently caused a variety of component shortages, and we were only able to acquire four unique lots of Varget for testing.   Our original plan was to test the four lots of Varget in .223 Remington with the 69 grain Nosler Custom Competition and in .308 Winchester with the 155 grain Hornady AMAX.

However, when shooting the 155 grain AMAX loads, the fourth shot of lot D caused the brass cartridge case to become stuck in the chamber, and the bolt handle sheared off when attempting extraction from the Rem 700 ADL. Consequently, that rifle was out of service for a time. Rather than waiting on the .308 Winchester to be repaired and completing the data collection on a different day, we decided to complete the study with a mid-weight bullet (53 grain Hornady VMAX) in .223 Remington. We had enough remaining powder in all four lots to load 40 more rounds of .223 Remington, but not to determine a safe powder charge and prepare new experimental loads in .308 Winchester. We will present and discuss the results of the four shots fired with each Varget lot in the .308 Winchester, along with the results of the complete data sets (ten shots with each lot of powder and bullet combination) with each of the two bullets tested in the .223 Remington, but we consider the results with the .223 Remington to be more definitive, because they are based on a full data set.

Feedback from the first study has often included questions about testing temperature variations. We remain unsatisfied with current temperature testing methods, especially the approach of conditioning ammunition at a given temperature (high, low, and room) and then firing in an ambient temperature rifle near room temperature. We've begin working with a temperature controlled box that should allow the entire rifle and cartridge to be held at the desired temperature and the shot triggered remotely. This design does not separate out powder effects from other effects that may also be temperature dependent, but it would inform the audience of the total expected velocity variation the system of rifle and cartridge together would give at high, low, and room temperatures. This is a more realistic experimental realization to the first shot fired under hunting conditions on a hot summer day, a cold winter day, and a moderate temperature day. While we work to perfect an experimental design for to quantify temperature dependence, we present our results for lot to lot variations in Varget.

**Method**

Powder was acquired from four different lots of Varget, designated A-D, and allowed to acclimate in the same storage area for over a year. All lots were kept in their original canisters and opened briefly from time to time. The temperature at which the powder was stored varied seasonally from 55 F in the winter to nearly 70 F in the summer. The relative humidity varied from 25% to 50%, which due to the high elevation (7000 ft) in Colorado, would be comparable to atmospheric moisture content of 18% to 36% relative humidity at sea level. Ten rounds were loaded with each of the four lots in each of the three test loads (120 rounds total). The 69 grain NCC bullets were loaded in Lake City brass using Fed 205M primers and 26 grains of Varget. The 53 grain VMAX loads were loaded in Lapua brass using Fed 205M primers and 27 grains of Varget. The 155 grain AMAX loads in .308 Winchester were loaded in Laupa brass with Fed 210M primers and 47 grains of Varget. Brass preparation method included cleaning in stainless tumbling media, reaming the primer pockets, and chamfering the flash hole and case neck with appropriate tools.

The two test rifles were a factory Remington 700 ADL chambered in .223 Remington (24" barrel) and a factor Remington 700 ADL chambered in .308 Winchester (22" barrel). Both rifles have twist rates of 1 turn in 12 inches. The longer bullets 69 grain stabilize fine in the relatively slow twist due to the thinner atmosphere at high plains and mountain elevations in Colorado. Before the experiment, the rifle barrels were cleaned thoroughly with our standard laboratory procedure. Velocities were measured with a CED Millenium chronograph with LED sky screens. Previous work has shown that the accuracy of these chronographs is about 0.3%. Four warm-up shots were fired to condition the bore and warm the barrel. Then one shot was fired from each lot of powder in sequence to interleave the lots of powder as the forty shots for record were fired. Data was recorded in a field notebook for later entry into a spreadsheet for analysis. Interleaving the shots prevented confounding effects from barrel friction and barrel temperature changing in time. The shots were not carefully timed, but a regular cadence was maintained. If a break was needed for some reason, between two and four additional warm-up shots were fired, depending on the length of the break, so the experiment would not resume with a cold bore.  In the actual firing sequences, breaks were only due to the occasional cease fire and only lasted a few minutes.

Care was taken to ensure experimental accuracy.  The same person loaded all the test rounds for each bullet and cartridge combination for all four powder lots, and all the rounds of each bullet and cartridge combination were loaded within a few days of each other and shot by the same person on the same day over the same chronograph.  The lots of brass, primers, and bullets were the same for each combination of cartridge and bullet, so that the only variable in the experiment was the lot of powder.  Interleaving shots from the different lots of powder exposed each different lot of powder to the same range of conditions as far as barrel heating and fouling are concerned.

After the data was entered into a spreadsheet, the average (mean) velocities and uncertainties in the means were computed for each of the four lots of powder. To present the spread of velocities clearly, velocity differences were computed relative to the lot which gave the slowest average velocity.  The uncertainties in the mean were computed as the standard error of the mean using the spreadsheet standard deviation function divided by the square root of the number of shots fired for each lot.

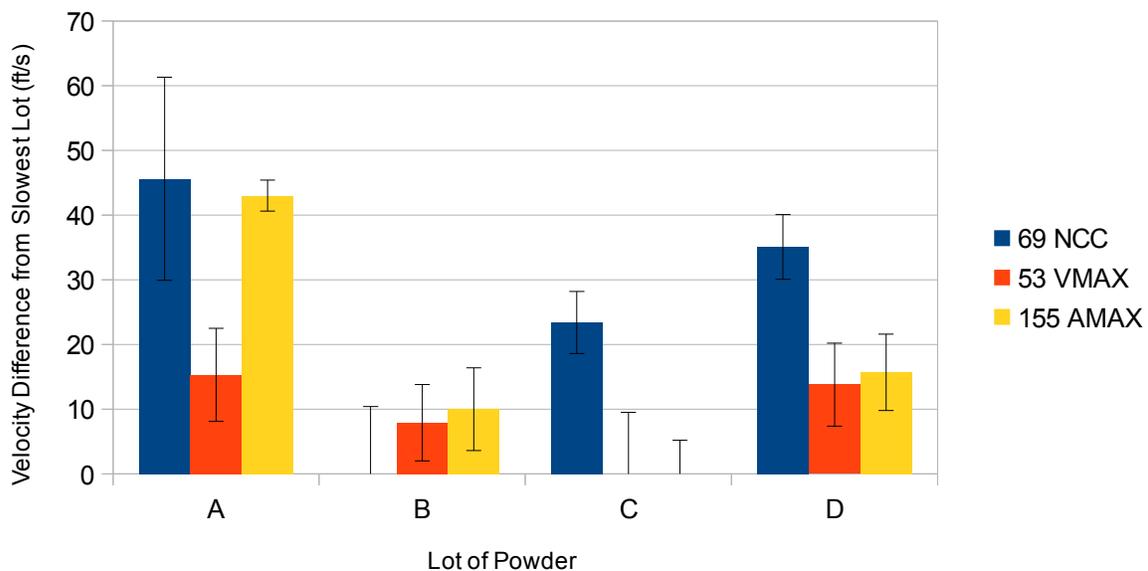

*Figure 1: Average velocity variation for the four lots of Varget in each of the three experimental loads.*

## Results

Variations in average velocity (and uncertainties) for the three loads tested are shown in Figure 1. Mean velocities for the three loads are shown in Table 1.  Lot B had the lowest velocity in .223 for the 69 grain NCC bullet, and Lot C had the lowest mean velocity in the other two loads. The mean velocity data is reported for the 155 grain AMAX in .308 Winchester, even though only four rounds were fired before the cartridge case became stuck in the chamber, presumably due to an overpressure condition. In spite of only firing four shots for each lot in the .308 Winchester, the uncertainties are below 7 ft/s because the shot to shot velocity variations were low within each lot. Ten shots were fired with each lot in each of the loads in .223 Remington.

| Load / Lot | A | B | C | D |
|---|---|---|---|---|
| .223 Rem 69 NCC | 2980.7 | 2935.1 | 2958.5 | 2970.2 |
| .223 Rem 53 VMAX | 3195.0 | 3187.6 | 3179.7 | 3193.5 |
| .308 Win 155 AMAX | 2784.0 | 2751.0 | 2741.0 | 2756.7 |

*Table 1: Average velocities (ft/s) for each lot of powder in each cartridge.*

**Discussion**

The velocity variations of the 53 grain VMAX load are well correlated with the 155 AMAX load with a correlation coefficient of 0.829. In contrast, velocity variations between the two .223 Remington loads is more modest with a correlation coefficient of 0.539. This suggests that the experiment is sensitive to lot to lot contributions of velocity variation, but due to the relatively small lot to lot variations with Varget, other experimental factors may also be contributing to the observed variations.

The date codes on the powder lots indicate that the four powder lots represent a range of years of manufacture from 2002 to 2011, which suggests Hodgdon has probably done a good job with quality control of Varget for many years. It is not clear why H4831 showed larger velocity variations with different lots of powder (> 100 ft/s, Courtney and Courtney 2013) than Varget. The specific lots of powder that were acquired for testing may be a factor, and we expect different variations may be observed from different lots of powder or with different loads.

We remain puzzled by the sticking of the cartridge case loaded with 47 grains of Varget (lot D) and the 155 AMAX, which was presumably due to an overpressure condition. We've carefully worked up many loads with Varget over the years, and our records indicate safe (no pressure signs) .308 Winchester loads with 47 grains of Varget and a variety of bullets including the 155 grain AMAX, the 150 grain Nosler E-Tip (a longer, harder solid copper bullet), the 155.5 grain Berger FullBore Boattail, the 168 grain Berger VLD, and the 125 grain Nosler Ballistic Tip. An average velocity under 2760 ft/s is not pushing the velocity threshold for a 155 grain bullet in a .308 Winchester with a 22" barrel, and the 155 grain AMAX is a relatively soft bullet, having a softer core than hunting bullets, and a thin, precision jacket. We can usually tell when a load is getting warm from flattened primers, case head expansion, and loose primer pockets. In 17 years of reloading, this was our first and only stuck case, demonstrating that experience is no guarantee against high pressure conditions.

One hypothesis for the stuck case with the 155 AMAX relates to using Lapua brass. Most of the loads mentioned above had been developed with Remington (R-P) brass, which we measured to have an average weight of 164.3 grains and an average case volume (once fired) of 56.8 grains of water. In contrast, the test loads in this experiment used Lapua brass, which we measured to have an average weight of 171.6 grains and an average case volume (once fired) of 56.2 grains of water. The fact that the Lapua cases are heavier (contain more brass) leaves less case capacity which causes higher peak pressure for a given powder charge. Entering the relevant quantities in QuickLoad V.3.6 suggests that the peak pressure with the Lapua cases is still safe at 57585 psi, but it also shows a peak pressure 1433 psi higher than expected for the case capacity of the R-P cartridge cases. QuickLoad also suggests a muzzle velocity of 2786 ft/s, close to the actual observations.

Clearly, reloading is not an exact science, and high pressures can occur due to factors that may not be predictable or easily explained. Understanding lot variations of components remains an ongoing challenge. At their web site, Hodgdon recommends working up new loads when the lot number of any component is changed, and it is evident that this should include a new lot of bullets, or brass cases, or primers. The need to work up loads and approach maximum loads carefully also includes new lots of Hodgdon Extreme powder (Hodgdon, 2012b):

*For all brands of powders use only the components shown. If the reloader makes any changes in components or gets new lot numbers, he should begin again with the starting loads and work up to maximum cautiously.*

**Acknowledgments**

This research was funded by BTG Research (www.btgresearch.org) and the United States Air Force Academy. The authors appreciate use of ranges at the Colorado Rifle Club and valuable input from Dr. Amy Courtney.

**References**

Hodgdon Powder Company. 2012a. Extreme Rifle Powders. Accessed 19 Nov 2012. http://www.hodgdon.com/smokeless/extreme/page2.php#top

Hodgdon Powder Company. 2012b. Reloading Data Center. Accessed 19 Nov 2012. http://data.hodgdon.com/main_menu.asp

Courtney, E.R. and Courtney M.W. 2013. Powder Lot Variations: A Case Study with H4831 – Hodgdon Extreme. Target Shooter Online. January 2013. pp. 80-83. See Also: http://www.dtic.mil/dtic/tr/fulltext/u2/a572333.pdf

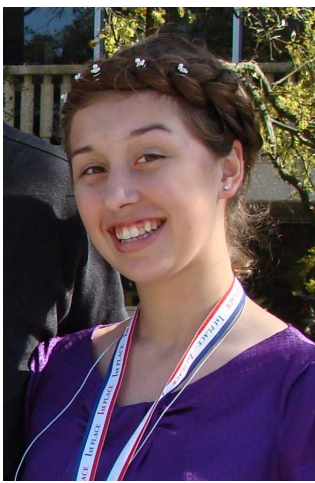

**About the Lead Author**
Elya Courtney is shown at the 2013 Louisiana Science and Engineering Fair. For the last several years, the Physics and Astronomy category has been dominated by projects in theoretical astrophysics. Against steep odds, Elya captured first place as a freshman with her project using ballistics to investigate whether the aerodynamic drag on supersonic projectiles is proportional to air density. Elya began serving as her dad's range assistant at 7 years of age, and has grown into both a talented scientific investigator and a skilled competitive shooter. Now 15, Elya is competitive in both long range precision rifle contests shooting metallic targets (prairie dogs) as well as paper targets in F-Class and similar matches. She has recently started shooting competitive bench rest matches to hone more fundamental loading and shooting technique independently of long range wind doping skills. Her "sweet shooter" rifle and load is a custom Lilja 6.5-284 barrel on a Remington 700 action shooting 140 grain Hornady AMAX bullets in Lapua cases with a max load of H1000 and Fed 210M primers.